\documentclass[12pt,column,A4paper]{article}

\usepackage[margin=1in]{geometry}
\usepackage{graphicx}
\usepackage{amsmath}
\usepackage{amssymb}
\usepackage{booktabs}
\usepackage{float}
\usepackage{indentfirst}

\newcommand{\chapquote}[3]{\begin{quotation} \textit{#1} \end{quotation} \begin{flushright} - #2\end{flushright} }

\usepackage[pagebackref,breaklinks,colorlinks]{hyperref}
\usepackage[nodisplayskipstretch]{setspace}
\newcommand\keywords[1]{\textbf{Keywords}: #1}

\usepackage[capitalize]{cleveref}
\crefname{section}{Sec.}{Secs.}
\Crefname{section}{Section}{Sections}
\Crefname{table}{Table}{Tables}
\crefname{table}{Tab.}{Tabs.}

\begin{document}

\title{Construct sparse portfolio with mutual fund's favourite stocks in China A share market}

\author{Ke Zhang\\
{\tt\small kezzhang@gmail.com}
}
\date{}
\maketitle

%%%%%%%%% ABSTRACT
\begin{abstract}
   Unlike developed market, some emerging markets are dominated by retail and unprofessional trading. China A share market is a good and fitting example in last 20 years. Meanwhile, lots of research show professional investor in China A share market continuously generate excess return compare with total market index. Specifically, this excess return mostly come from stock selectivity ability instead of market timing. However for some reason such as fund capacity limit, fund manager change or market regional switch, it is very hard to find a fund could continuously beat market. Therefore, in order to get excess return from mutual fund industry, we use quantitative way to build the sparse portfolio that take advantage of favorite stocks by mutual fund in China A market. Firstly we do the analysis about favourite stocks by mutual fund and compare the different method to construct our portfolio. Then we build a sparse stock portfolio with constraint on both individual stock and industry exposure using portfolio optimizer to closely track the partial equity funds index 930950.CSI with median 0.985 correlation. This problem is much more difficult than tracking full information index or traditional ETF as higher turnover of mutual fund, just first 10 holding of mutual fund available and fund report updated quarterly with 15 days delay. Finally we build another low risk and balanced sparse portfolio that consistently outperform benchmark 930950.CSI.

\end{abstract}
%%%%%%%%% BODY TEXT
\section{Introduction}

Active investing compare with passive investing is a popular topic during recent years. In developed country, lots of research shows active investing in average can't beat passive investing in decades\cite{French2008cost}\cite{Bogle2017common}\cite{Nanigian2019Performance}. The main reason they conclude are: active management fee is much higher than pass management fee. Active investing tax is less efficient than passive. Top active fund can't continuously stay in top performance category. Higher cost from higher turnover rate of active fund. Top fund AUM grow quite fast and Alpha decreased correspondingly. Most of these research focus on developed market like United States or Europe which are dominated by institution investor. Because of this, it is very hard to get excess return and Alpha especially calculate after total fee. On the other hand, in some emerge stock market like China, it seems active fund have better performance consistently even we compared after fee and tax. Most reason cause this is the market inefficiency due to short history of stock market and high percentage of retail trading.

Unlike developed stock market, history of China A share market is just 30 years. Shanghai Stock Exchange built at 1990/11/26. Shenzhen Stock Exchange built at 1990/12/01. In 1998, first 6 mutual fund company built and 5 mutual fund IPO. Number of mutual funds in this market increased from 0 to more than 10 thousand from 1998 to 2023Q1. The market capitalization hold by mutual funds arrived one trillion CNY at 2007Q1 firstly, achieved twenty five trillion CNY at end of 2022 which is 4th largest in world.\cite{imynsimy4655459}

However, institution investors are still a very small portion of this market. Until 2022Q3, institution investors just hold about 17\% of stocks measured by market capitalization. And mutual fund hold about 8\% market capitalization of total stock market.\cite{Li2022Astock}  

\begin{figure}[H]
\centering
  \includegraphics[width=1.\linewidth]{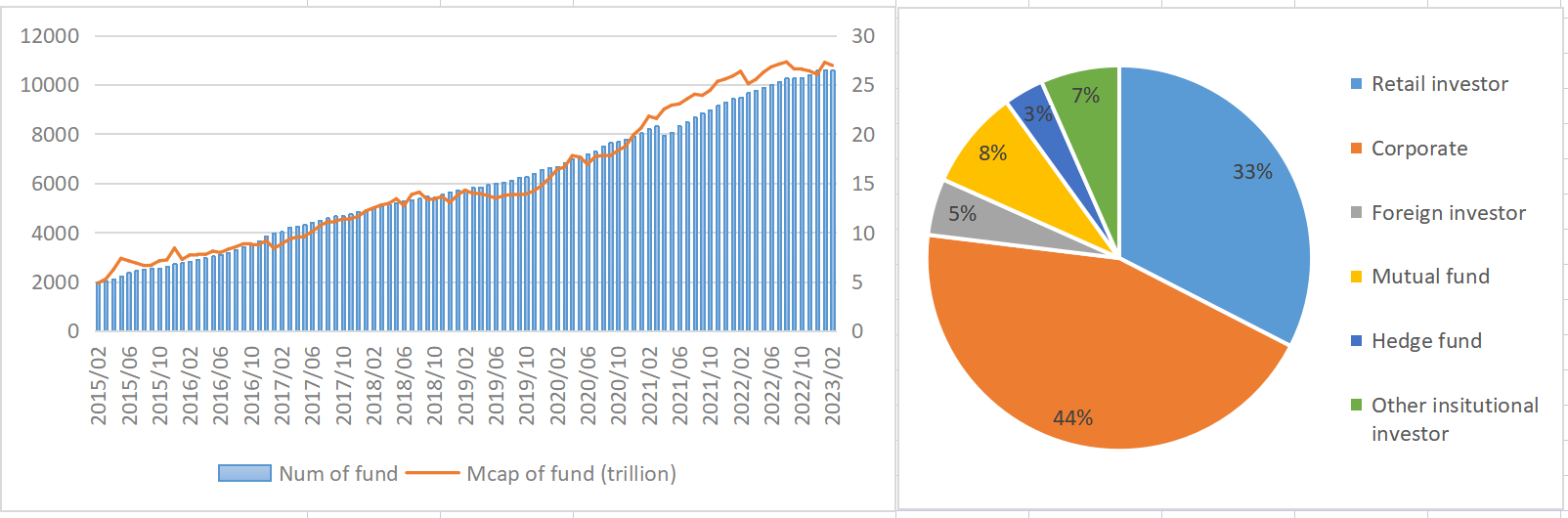}
   \caption{Size of mutual fund in history and investor distribution in China A share market}
   \label{fig:portion2}
\end{figure}

As institution investors market capitalization is just small part of China A share market, we expect institution investors such as mutual fund could generate excess return. Indeed lots of researchers show mutual funds in China A share market do get excess return or Alpha compare with total market return \cite{Chi2022Perform}\cite{Wool2021China}\cite{zhu2017lianghua}\cite{Zheng2015Herding}\cite{Hsu20220mutual}\cite{Tauni2017Emerging}. 

If mutual funds in China could beat market in average, how could we take advantage of these information. Researchers already shows top mutual fund can't stay in top category all the time. Sometimes worst fund last year will have better performance later as mean reversion. China largest index provider, China Securities Index Co., Ltd. (CSI), construct a partial equity funds index 930950.CSI which reflect the average performance of partial equity funds in China A share market. However, in 2023 Q1, 930950.CSI hold more than 5000 funds which is impossible to directly invest.

In this paper, our first goal is mimic 930950.CSI with a small stock universe using constraint sparse convex optimizer. As 930950.CSI is a long-only index, therefore our portfolio is also long-only which automatically inherit the L1-norm penalty which we will explain in detail later. Traditional index track method is minimizing the MSE of different between portfolio return and index return \cite{Brodie2009Sparse}\cite{Benidis2017sparse}. As our index holding information has time delay and just partial public information available. Thus it is much more difficult to track than full information index or traditional ETF. Therefore, we also add the industry exposure constraint in our portfolio. At last we will construct another more balanced and lower volatility portfolio which showed consistently beat the 930950.CSI. 

\section{Choose benchmark}
In this section, we will choose a benchmark which could represent the average performance of mutual fund in China A share market. 

China Securities Index Co., Ltd. (CSI), a financial market index provider jointly funded by the Shanghai Stock Exchange and the Shenzhen Stock Exchange in August 2005. It is Chinese largest index company which public more than 95\% of ETFs in China A share market. 

CSI mutual fund indices provide by CSI reflect the overall performance of all CSRC regulated mutual funds and their sub-classes. The indices provide benchmarks and underlyings for fund investors. 

CSI partial equity funds index (930950.CSI) is one of CSI mutual fund indices which provide benchmarks for partial equity Funds. It includes all funds that open public more than 3 months and whose lower limit of investment scope is 60\% and weight by fund shares which have larger capacity than equal weight method. Until 2023 Q1, there are more than 30 funds or fund of funds using 930950.CSI as benchmark.

\begin{table}[H]
\centering
\begin{tabular}{||c c c||} 
 \hline
Year & Average of partial equity funds
 & 930950.CSI \\ [1ex] 
 \hline\hline
 2022 & -20.68 & -21.80  \\ 
 \hline
 2021 & 7.43 & 4.05 \\
 \hline
 2020 & 5.61 &	5.15  \\
 \hline
 2019 & 43.07 &	43.74  \\
 \hline
 2018 & -20.45 & -24.58  \\ 
 \hline
 2017 & 12.43 &	12.63  \\ 
  \hline
 2016 & -12.88 & -17.00   \\ 
\hline
 2015 & 47.91 &	37.52 \\  [1ex] 
 \hline
\end{tabular}
\caption{Average return of partial equity funds vs return of 930950.CSI, Data from fund.eastmoney.com and uqer.datayes.com}
\label{tab:t1}
\end{table}

In \cref{tab:t1}, we compare average return of partial equity funds and 930950.CSI which they are highly correlated. Meanwhile 930950.CSI performance is slightly worse than average return of Partial Equity Funds. We think the main reason is 930950.CSI weighted by shares of funds that overweight large AUM funds. Also 930950.CSI exclude small AUM funds public less than 3 month. As we known, equity mutual fund performance is negative correlated with size of AUM and higher IPO return of small AUM funds.  

After we decide our benchmark, let's compare this partial equity funds index with other popular passive index in China A share market. 000300.CSI index consists of the 300 largest and most liquid A-share stocks which aims to reflect the overall performance of large cap stock in China A-share market. 000905.CSI index consists of the 500 largest and most liquid A-share stocks without stocks in 000300.CSI universe which aims to reflect the overall performance of mid-small cap stock in China A-share market. 930903.CSI is total market index in China A-share market.

\begin{table}[H]
\centering
\begin{tabular}{||c c c c||} 
 \hline
 Index & Total Return(\%) & Vol & Sharpe \\ [0.8ex] 
 \hline
 930950.CSI & 60.49 & 23.52 & 0.139 \\ 
 \hline
 000300.CSI & -23.36 & 25.97 & -0.068 \\
 \hline
 000905.CSI & 24.95 & 29.46 & 0.06 \\
 \hline
 930903.CSI & 3.21 & 26.42 & 0.012 \\[1ex] 
\hline
\end{tabular}
\caption{Compare different indices from 2007/12/31 to 2023/03/01, data from uqer.datayes.com}
\label{tab:t2}
\end{table}
In \cref{tab:t2}, Here we compare performance of different indices with total return, annualized volatility ( using log return), Sharpe ratio ( using log return). Not surprise, partial equity funds index outperform other indices with higher total return and lower volatility. Which means, we pick the best perform index in China A share market as benchmark to track and beat.

\section{Trade setting and assumption}
Firstly, as fund reports we used are quarterly base, trade time of our model set at 16 days after each quarter when last quarterly report firstly available. We assume we can just trade stock after it IPO 3 months later. We can't trade suspension stock. We can't buy stock at limit up and sell stock at limit down.

We assume commission cost is 5 bps for buy and 1.5 bps for sell. Our order assume trade-able stocks be filled fully with 1 cent slippage cost for each side. Our beginning cash is 100 million CNY which is average AUM of small mutual funds. Finally we assume we can just trade shares that can be divided by 100.

\section{Fund and stock screener}
In order to construct partial mutual fund portfolio, we need generate our stock universe firstly. As our data source doesn't include information about partial equity mutual fund, we need manually screen our funds. Our target funds are partial equity mutual funds listed more than 3 months.

\begin{itemize}
\item Only include Open-ended Funds( No Close-ended Funds).
\item Only include funds with categories Equity or Hybrid.
\item As we just care about active fund, so we exclude ETF, Enhanced ETF or ETF connected fund.
\item Exclude QDII, FoF, Structured Fund or Guarantee fund.
\item Get rid of duplicated fund, only include one fund if several funds with same name but end with different alphabet such as A,B,C,D
\item Exclude fund listed less than 3 months.
\end{itemize}

As universe of our portfolio are stocks. Then we filter stocks from our candidate funds. 
\begin{itemize}
\item At 16 days after each end of quarter (earliest time to get quarterly report), we get first 10 holding from candidate funds which are only holding information available from quarterly fund report.
\item Only include stocks in A-share market ( No B-share, H-share or ADR and so on).
\item Only include stocks listed more than 3 months.
\item Only include stocks is trade-able.
\end{itemize}
Here we only use quarterly report information as semi-annual and annual fund report delay too much. We doesn't exclude ST or ST* stocks as we thought these stocks selected by mutual funds still create Alpha and are liquid enough.

After we get our stock universe, now we want to choose a way to weight our stocks. 930950.CSI use fund share as weight for their index. Here we construct three weight methods and compare their performance. Firstly we use market capitalization of holding stocks which is closed to the weight method by 930950.CSI. Also we create a method that weight stocks by how many funds hold it. Finally, we construct third method which use sum of weight of stocks in each funds.

\begin{equation}
  w_{i} = \frac{\sum_j{mcap_{i,j}}}{\sum_i\sum_j{mcap_{i,j}}}
\label{eq:e1}
\end{equation}
$mcap_{i,j}$ is market capitalization of stock i hold in fund j.\\
\cref{eq:e1} is weight of stock by market capitalization of holding in mutual funds. 

\begin{equation}
  w_{i} = \frac{\sum_j{I_{i,j}}}{\sum_i\sum_j{I_{i,j}}}
\label{eq:e2}
\end{equation}
$I_{i,j} = 1$ if stock i in fund j else 0.\\
\cref{eq:e2} is weight of stock by numbers of holding in mutual funds.

\begin{equation}
  w_{i} = \frac{\sum_j{p_{i,j}}}{\sum_i\sum_j{p_{i,j}}}
\label{eq:e3}
\end{equation}
$p_{i,j}$ is weight of stock i in fund j .\\
\cref{eq:e3} is weight of stock by weight sum of holding in mutual funds.

\cref{eq:e1} consider the effect of both market capitalization and number of holding. \cref{eq:e2} only consider the effect of the number of holding. \cref{eq:e3} consider the effect of both weight of stock in mutual funds and number of holding in mutual funds. Then let's look at the performance of each method.

\begin{table}[H]
\centering
\begin{tabular}{||c c c c||} 
 \hline
 Method & Total Return(\%) & Vol & Sharpe \\ [0.8ex] 
 \hline
 Market capitalization of holding & 108.47  & 25.53 & 0.326 \\ 
 \hline
 Numbers of holding & 110.37 & 24.56 & 0.343 \\
 \hline
 Weighted sum of holding  & 112.01 & 25.10 & 0.338 \\
\hline
\end{tabular}
\caption{Performance of different weight method from 2013/12/31 to 2023/03/01, data from uqer.datayes.com}
\label{tab:t3}
\end{table}
In \cref{tab:t3}, we weighted stocks by each method mentioned above. and constructed a simple long only portfolio. Here volatility and Sharpe ratio are calculated using log return. As we can see market capitalization weight method has worst return and volatility. Weighted sum of holding has highest return but volatility is slightly worse than weighted by numbers of holding. In the section below, when we want to track index closely, we choose the market capitalization weight method similar with index which is fund share weight. On the other hand, when we want to build a portfolio beat the index, we select weighted sum of holding with best performance.

\section{Sparse convex optimization with constraint}

\chapquote{"Diversification is the only free lunch in investing."}{Harry Markowitz}

As original partial equity mutual fund index 930950.CSI hold thousands of funds each quarter which is impossible to directly invest. We want to build a much smaller stock universe portfolio to track or beat it. Convex optimization with L1-norm fit our requirement.   

Portfolio optimization is a very important part of investing. Harry Markowitz provide mean-variance optimization in 1950s\cite{markowitz1952portfolio}. Steven Boyd publish convex optimization book at 2004 \cite{Boyd2004convex}. Also people realized we can transfer some none convex problem to convex problem \cite{Boyd2004convex}. For example we can transfer L1-norm optimization problem to equivalent quadratic optimization problem.

\begin{equation}
\begin{aligned}
\min_{w} \quad & ||A^T w-B||^2_2+\lambda ||w||_1\\
\textrm{s.t.} \quad & w \in \mathbb{C}\\
\label{eq:e4}
\end{aligned}
\end{equation}
Here $||w||_1 = \sum_i{|w_i|}$.\\
\cref{eq:e4} is convex optimization problem with L1-norm.

This problem is equivalent to solve:

\begin{equation}
\begin{aligned}
\min_{w} \quad & ||A^T w-B||^2_2+\lambda 1^Tv\\
\textrm{s.t.} \quad & w \in \mathbb{C}\\
&-v \leq w\leq v    \\
\label{eq:e5}
\end{aligned}
\end{equation}
\cref{eq:e5} is a convex quadratic optimization problem which is also another form of \cref{eq:e4}.

Optimization with L1-norm penalty has been showed several properties:
\begin{itemize}
\item Bring in the sparsity. This is one of most important reasons why people prefer L1-norm penalty. Because of sparsity, it could add explanation to model parameters. In portfolio selection, people consider use L1-norm to reduce number of holding or number of trading. 
\item Long-only portfolio optimization naturally include a L1-norm penalty.  
When we put two constraint $w_i \geq 0$ and $\sum_i{w_i} = 1$ into our problem. Then naturally we have $||w||_1 = \sum_i{|w_i|} = \sum_i{w_i} = 1$ which is a fixed number independent with w. Thus there are no more work we need to do to put sparsity requirement to long-only portfolio optimization.
\item Similar with L2-norm penalty, L1-norm penalty could stabilizes the estimators which usually used to avoid singularity or ill-condition of problem.\cite{Brodie2009Sparse}
\end{itemize} 

\section{Construct sparse stock portfolio to track 930950.CSI}

Build a portfolio to track full information index or ETF is well-known and there are lots of research and application about it. However, mutual fund just public their topic 10 holdings every quarter and will delay 15 days which make our problem much more difficult than traditional tracking goal.

As we mentioned before, 930950.CSI have better return and lower volatility than other passive benchmark index. However, 930950.CSI hold thousands of funds each period which is impossible to invest directly. Thus in this section, we will construct a sparse stock portfolio to track the 930950.CSI closely.

Here we construct our portfolio by this way:
\begin{equation}
\begin{aligned}
\min_{w} \quad & - \alpha^T w + \beta w^T\Sigma w + \kappa||R_{stock}^Tw-r_{bench}||^2_2+ ||w||_1\\
\textrm{s.t.} \quad &0 \leq w \leq \gamma \\
&\phi_1 \leq Aw \leq \phi_2 \\
& \sum_i{w_i} = 1 \\
\label{eq:e6}
\end{aligned}
\end{equation}

First of all, as weight method of 930950.CSI is closed to market capitalization, we rank stock by market capitalization of holdings and take first half of them as our portfolio universe which represent the stocks that favorite by mutual funds.

Here $\alpha$ is Alpha of each stock. We still use weight cap of holdings as Alpha to replicate 930950.CSI. Then we standardize our Alpha. After that we winsorize standardized Alpha by 3 standard deviation from each side. 

$\beta$ is risk parameter to control how important of our risk model are. $\Sigma$ is our covariance model to control risk of portfolio. We use sample covariance by rolling two year log return of stocks. In order to avoid covariance singularity or ill-condition we shrink our covariance by method mentioned in \cite{Chen2010Shrink}.

$||R_{stock}^Tw-r_{bench}||^2_2$ is a common tracking measure called empirical tracking error.\cite{Benidis2017sparse} We minimize it to closely track our benchmark. $R_{stock}$ is return of stocks in our universe. $r_{bench}$ is return of index. Here we use rolling 3 month data to do this mean square error minimization because fund report update quarterly.

$||w||_1$ bring sparsity into our portfolio. As we also have constraint $w_i \geq 0$ and $\sum_i{w_i} = 1$ , thus $||w||_1 = \sum_i{|w_i|} = \sum_i{w_i} = 1$. This term is a constant that independent with w.

As law requirement, Mutual fund in China are long-only with 10\% single stock constraint. Thus we induce $0 \leq w \leq \gamma$. this inequality make sure we are long-only portfolio and weight of each stock less or equal than $\gamma$, in order to track index closely, here we choose $\gamma$ equal to 0.01 that maximum hold position of each stock is less than 1\%.

Just try to track index from individual stocks is not enough especially our portfolio is sparse and our benchmark is not an ETF. Our idea is also track the industry exposure of benchmark quarterly. As we don't have whole history holding of 930950.CSI. Here we use $\phi$ as industry exposure of market capitalization weight of fund stocks which is similar with weight method of 930950.CSI. In order to make portfolio feasible, we use inequality constraint instead of equality constraint. A is a $M \times N$ matrix with information if stock i in industry j where $i\leq M,j\leq N$.

Finally, $\sum_i{w_i} = 1$ make sure w is percentage weight of portfolio.  

\begin{figure}[H]
\centering
  \includegraphics[width=.8\linewidth]{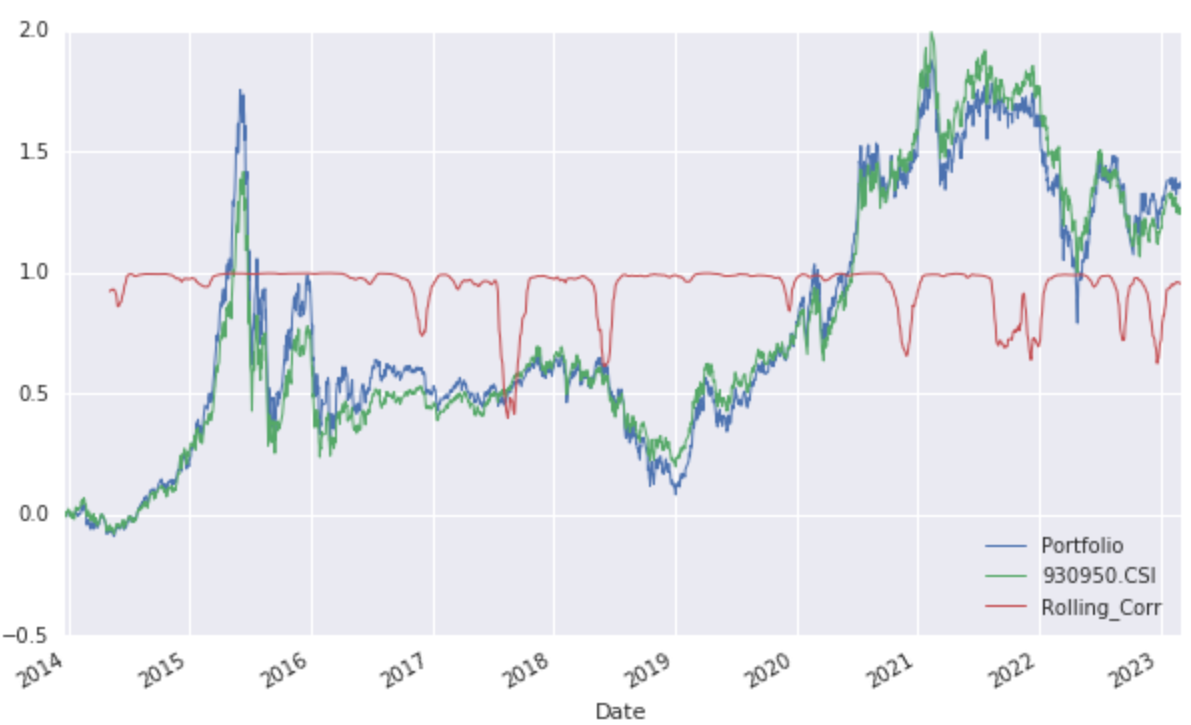}
   \caption{Return and correlation between Portfolio and 930950.CSI}
   \label{fig:track_graph}
\end{figure}

In \cref{fig:track_graph}, we successfully track the 930950.CSI closely, the median 3 month rolling correlation between our portfolio and index is as high as 0.985. We could find most time the correlation is very close except some short period. We thought the main reason are during that period, lots of mutual fund change their position a lot and we can just saw the first 10 holdings of mutual fund quarterly and report delayed 15 days. Therefore it is much more difficult to track than full information index or ETF. Some researcher already show the Herding behavior in institutional investors \cite{Zheng2015Herding}.

\begin{figure}[H]
\centering
  \includegraphics[width=.9\linewidth]{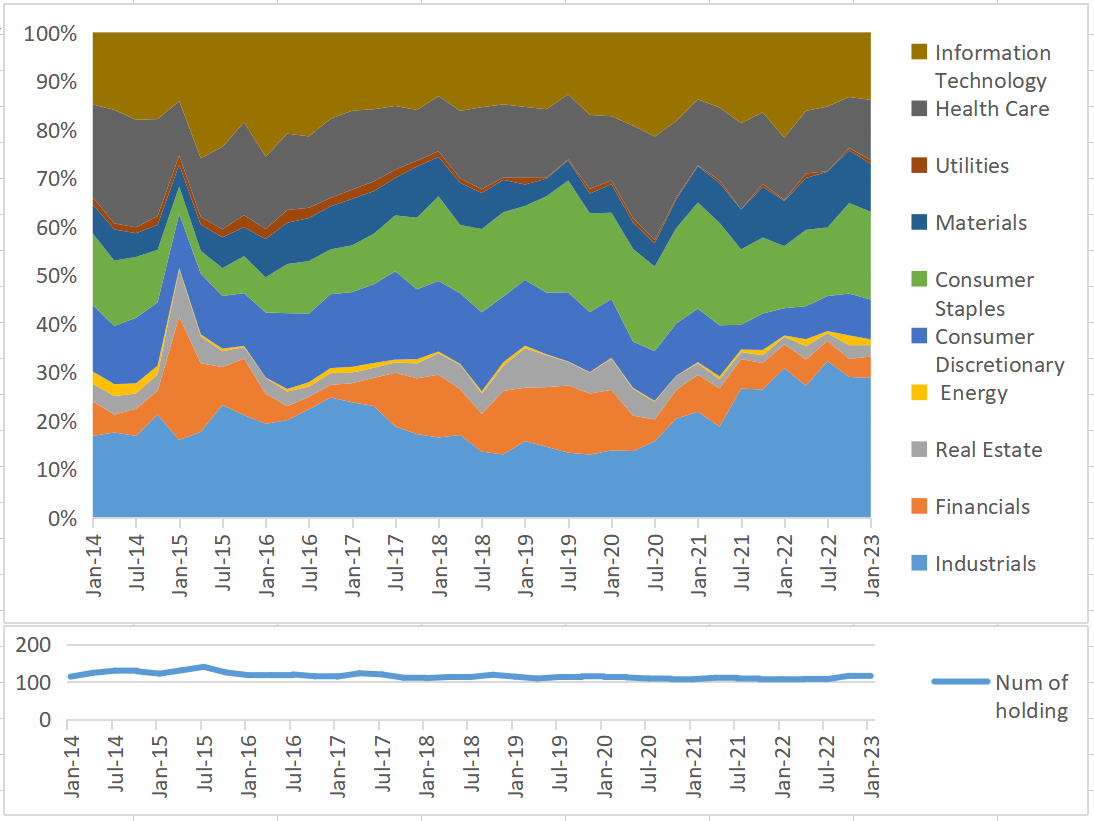}
   \caption{Industry exposure of portfolio and number of holdings}
   \label{fig:track_indu}
\end{figure}

In graph above, we showed the industry exposure of our portfolio which is optimized closely to 930950.CSI. We use CSI Industry Classification Standard Level I to get stock industry exposure. Communication Services sector is almost 0\% exposure all the time so we doesn't show on graph above. We can find sector Industry, Information Technology, Health care and Consumer Staple are over weighted that more than 10\%.

Meanwhile, we showed number of holdings in our portfolio. As we have a 1\% maximum holding limit, we have to hold more than 100 stocks each period. From graph we see that our holding is slightly more than 100 each period (about 120 in average and 25\% of universe). Thanks to the sparsity property of L1-norm penalty in portfolio optimizer. 

\section{Construct sparse portfolio beat 930950.CSI}

Here we construct another portfolio with optimizer but different setting:
\begin{equation}
\begin{aligned}
\min_{w} \quad & - \alpha^T w + \beta w^T\Sigma w + ||w||_1\\
\textrm{s.t.} \quad &0 \leq w \leq \gamma \\
&Aw \leq \phi \\
& \sum_i{w_i} = 1 \\
\label{eq:e7}
\end{aligned}
\end{equation}

Here we create another portfolio which will beat partial equity funds index, 930950.CSI. 

First of all, we choose the weight method with weight sum of mutual fund holding which showed higher return than other two weight method. Then we still select top 50\% of stocks as the favorites by mutual fund industry. 

Also, we know China A share market has very high volatility and lots of paper showed select stock by low volatility in China A share market could improve not only annual return but also Sharpe ratio. Therefore we sub-select our universe by top 100 lowest max-flat weighted volatility stocks. \cite{Zhang2023Adjust}\cite{Blitz2021the}\cite{Jansen2021Ano}.  

As our goal is beat 930950.CSI instead of track it closely and we have a much smaller universe after sub-select universe. We relax the single stock constraint $\gamma$ from 1\% to 10\% which is same as the limit of single holding of mutual fund. 

Finally, we restrict $\phi$ each industry exposure of our portfolio within 10\% to reduce the sector risk and build a more balanced portfolio.

\begin{figure}[H]
\centering
  \includegraphics[width=.8\linewidth]{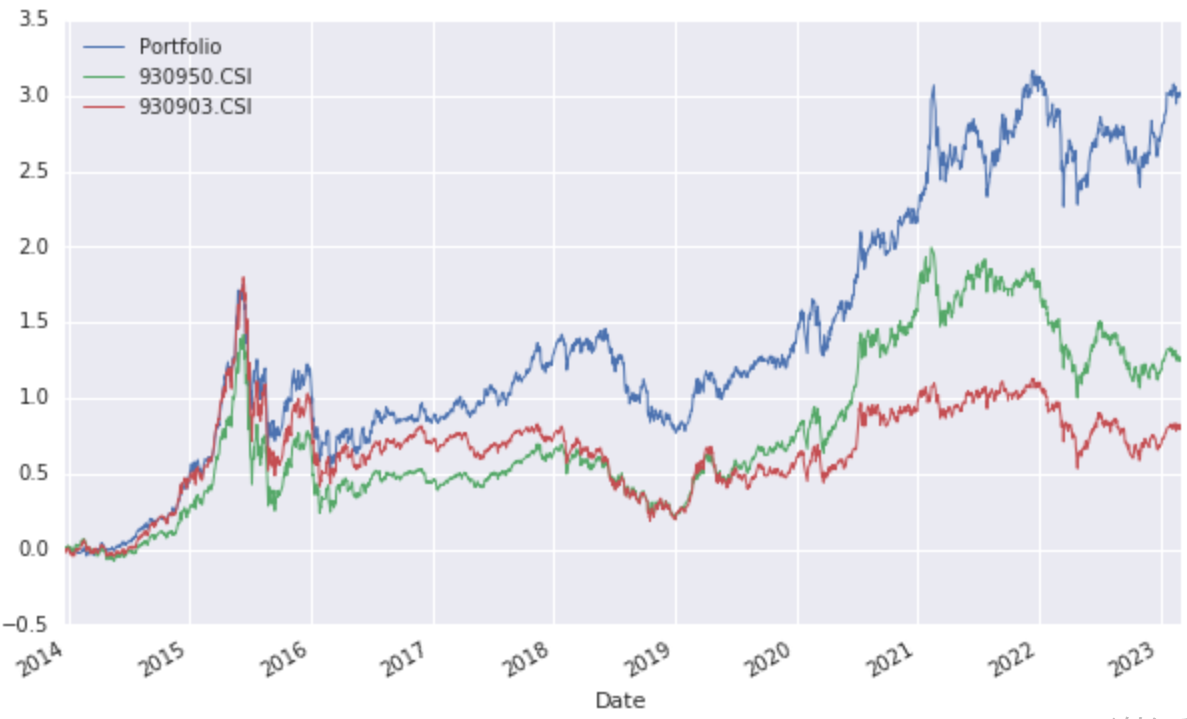}
   \caption{Performance of portfolio compare with benchmark}
   \label{fig:beat1}
\end{figure}

\begin{table}[H]
\centering
\begin{tabular}{||c c c c||} 
 \hline
 Methods & Total Return(\%) & Vol & Sharpe \\ [0.8ex] 
 \hline
 Portfolio & 302.27  & 22.22 & 0.71 \\ 
 \hline
 930950.CSI & 127.11 & 23.46 & 0.40 \\
 \hline
 930903.CSI  & 82.45 & 23.45 & 0.30 \\
\hline
\end{tabular}
\caption{Performance of portfolio compare with benchmark from 2013/12/31 to 2023/03/01, data from uqer.datayes.com}
\label{tab:t6}
\end{table}
Here volatility and Sharpe ratio are calculated by log return. From table above, obviously we can find our portfolio has better total return, lower volatility and higher Sharpe ratio compared with either partial equity funds index or total market index.

\begin{figure}[H]
\centering
  \includegraphics[width=1.\linewidth]{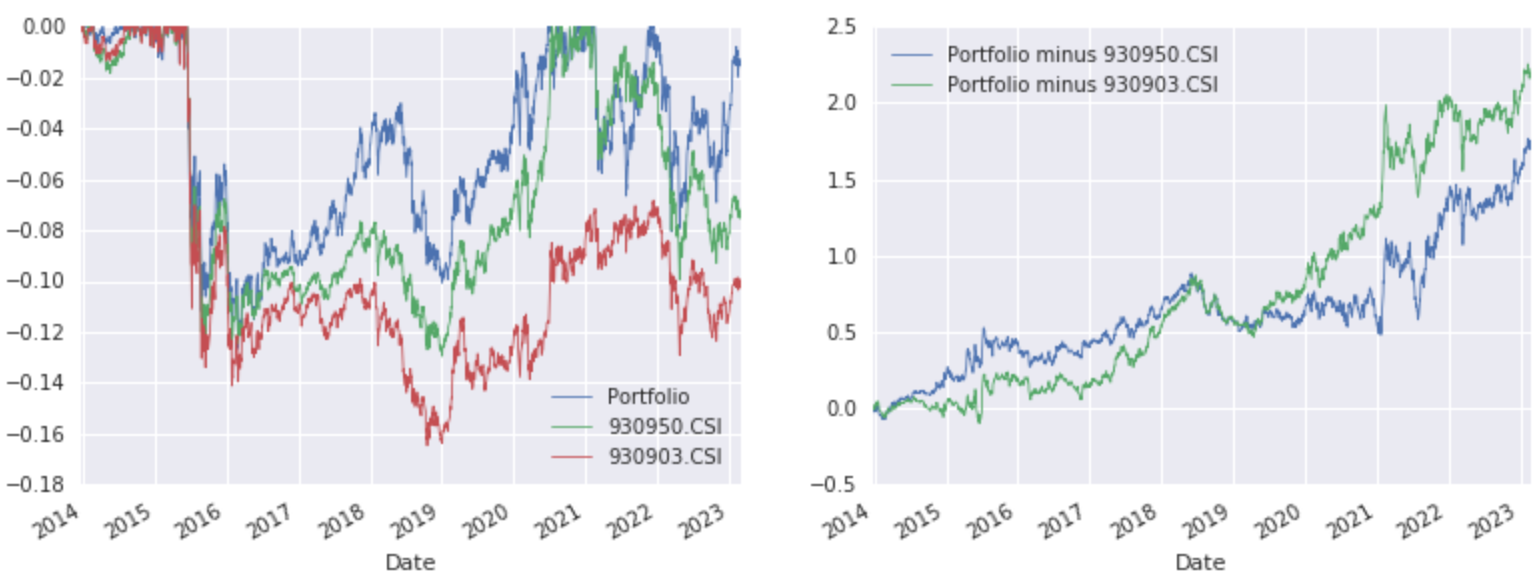}
   \caption{Under water graph and excess return graph}
   \label{fig:beat2}
\end{figure}

Meanwhile, we can compare the maximum drawdown of three strategies. Obviously, Our portfolio have lower maximum drawdown than 930950.CSI and 930903.CSI. 

Then let's consider use portfolio return minus 930950.CSI and portfolio return minus 930903.CSI. What we get two equity curve gradually go up which means our portfolio could beat this two index in long term. 

\begin{figure}[H]
\centering
  \includegraphics[width=.8\linewidth]{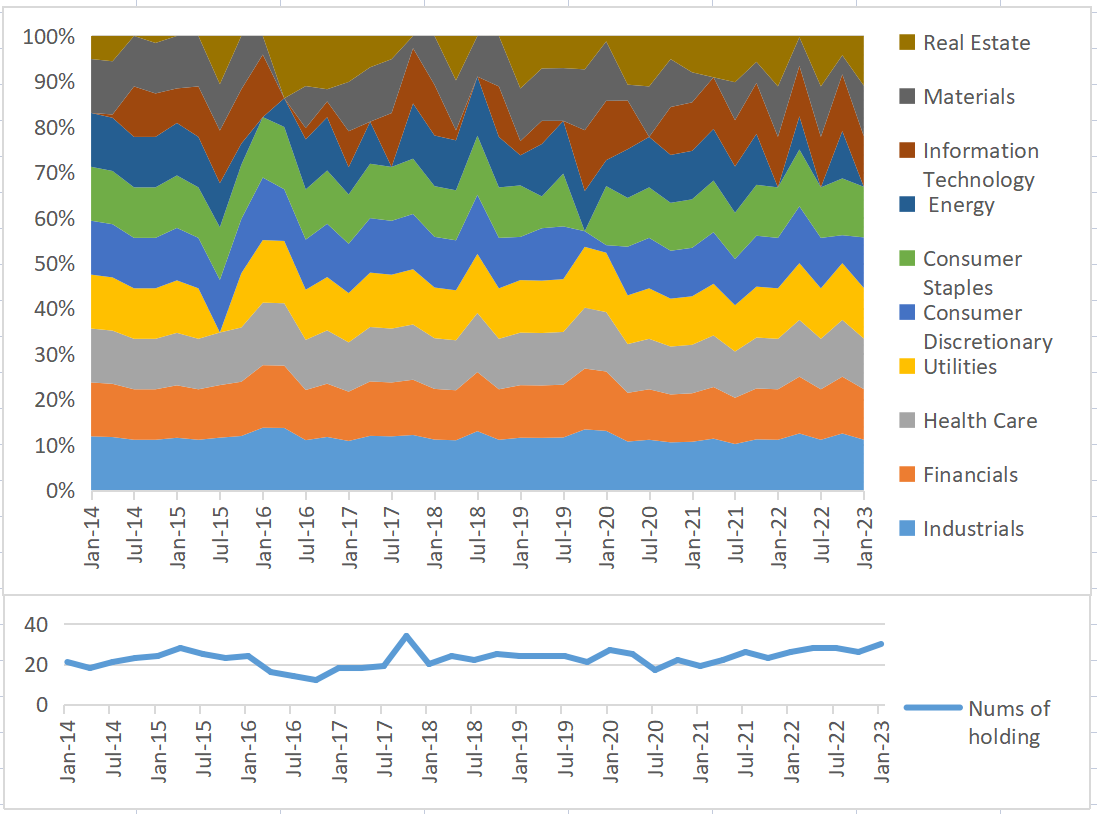}
   \caption{Industry exposure of portfolio and number of holdings}
   \label{fig:balance_indu}
\end{figure}

From graph above, all the industry exposure are restricted within 10\%. While Industry, Finance, Health care are all time about 10\% but lower than their weight in 930950.CSI. Energy, Utility obviously overweight compared with 930950.CSI. 

Meanwhile, our portfolio average holding is about average 25 stocks that is very sparse compare with 100 stock universe.

\section{Conclusion}
In this paper, we firstly do some analysis about mutual funds and mutual fund's holdings in China A share market. Similar with conclusion of other researchers, in last 20 years, mutual funds in China A share market outperform the total market index as the inefficiency of emerging market. Then we build the sparse portfolio with individual stock and industry exposure constraint to track partial equity  fund index. We get a high correlation (0.985) sparse portfolio. Finally, we build a low-risk balanced sparse portfolio that beat the partial equity funds index regard of return and Sharpe ratio consistently.  
%%%%%%%%% 

\end{document}